\documentclass{article}
\usepackage{spconf,amsmath,graphicx}
\usepackage{hyperref}
\title{An $l_{2}$-normalized spatial attention network for accurate and fast classification of 
brain tumors in 2D T1-weighted CE-MRI images
}
\name{Grace Billingsley\textsuperscript{1}, Julia Dietlmeier\textsuperscript{2}, Vivek Narayanaswamy\textsuperscript{1}, Andreas Spanias\textsuperscript{1}, Noel E. O'Connor\textsuperscript{2} }
\address{\textsuperscript{1} SenSIP Center, Arizona State University, Tempe, USA\\ \textsuperscript{2} Insight SFI Research Centre for Data Analytics, Dublin City University, Ireland}
\usepackage{xcolor}

\begin{document}

\mbox{}

\newpage
%\lipsum
%\onecolumn
\textbf{IEEE Copyright Notice}

\vspace{10pt}
© 2023 IEEE. Personal use of this material is permitted. Permission from IEEE must be obtained for all other uses, in any current or future media, including reprinting/republishing this material for advertising or promotional purposes, creating new collective works, for resale or redistribution to servers or lists, or reuse of any copyrighted component of this work in other works.

\vspace{10pt}
Accepted to be published in: IEEE International Conference on Image Processing (ICIP), Kuala Lumpur October 8-11, 2023.

%\ninept
%
\maketitle
\begin{abstract}
We propose an accurate and fast classification network for classification of 
brain tumors in MRI images that outperforms
all lightweight methods investigated in terms of accuracy.
We test our model on a challenging 2D T1-weighted CE-MRI dataset containing three types of brain tumors: Meningioma, Glioma and Pituitary. We introduce an $l_{2}$-normalized spatial attention mechanism that acts as a regularizer against  overfitting during training.
We compare our results against the state-of-the-art on this dataset and show that by integrating $l_{2}$-normalized spatial attention into a baseline network we achieve a performance gain of 1.79 percentage points. Even better accuracy can be attained by combining our model in an ensemble with the pretrained VGG16 at the expense of execution speed. Our code is publicly available at \url{https://github.com/juliadietlmeier/MRI\_image\_classification}.

\end{abstract}
\begin{keywords}
Deep learning, MRI image analysis, attention mechanisms, brain tumor classification
\end{keywords}
\section{Introduction}
\label{sec:intro}

Worldwide, cancer of the brain and other nervous system is the 10\textsuperscript{th} leading cause of death for adults \cite{CNN_2022_2}. Having their origin at the cellular level, brain tumors are one of the most lethal cancer types. Upon diagnosis, it is estimated that the median survival time is  14-18 months despite surgery, radiotherapy and chemotherapy treatments \cite{Cancer_2020}. Generally, brain tumors are diagnosed by means of non-invasive Magnetic Resonance Imaging (MRI), and high-sensitivity detection is attained using standard T1- and T2-weighted MRI. Long-term survival rates of cancer patients depend directly on  early diagnosis and accurate assessment. To support complementary therapies, early stage MRI image analysis provides both pre-operative and post-operative insights into a brain tumor's growth. Present clinical practice is to perform classification of cancerous tissue by biopsy prior to the surgery \cite{CNN_2020}. If the MRI images are available, the classification is performed by an experienced radiologist. However, manual classification and segmentation of brain tumors in images is time-consuming, expensive and may also be erroneous and subjective in certain complex cases. AI-based methods can help address these challenges and have the potential to become a key tool in assisting clinicians in the decision-making process. In particular, deep learning based approaches produce state-of-the-art results on medical image classification tasks in imaging modalities such as X-ray, CT and MRI. 
\begin{figure}[h]
  \centering
  \vspace{-10pt}
  \includegraphics[width=7cm]{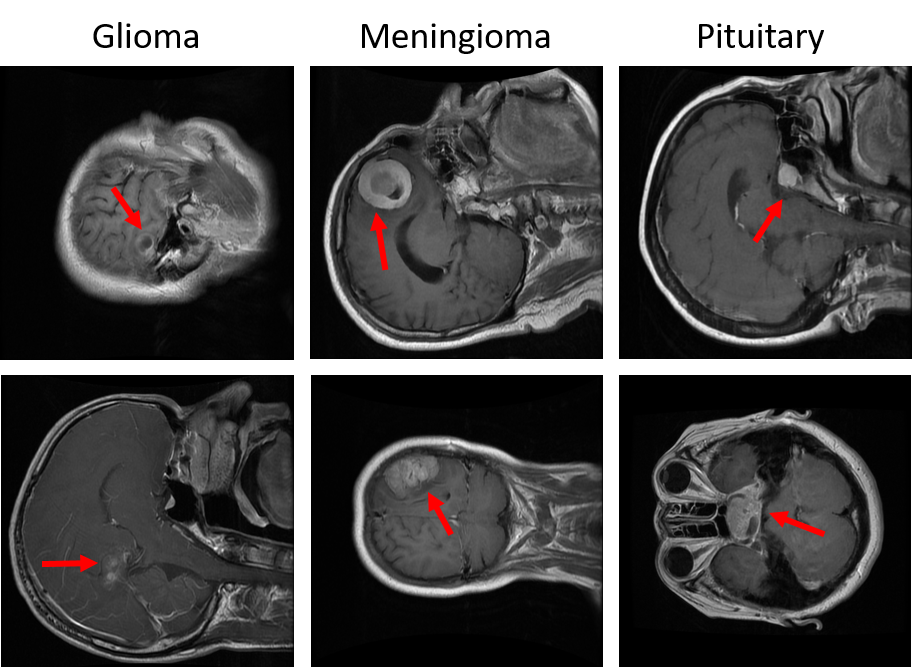} 
  \caption{Random sample from the 2D T1-weighted CE-MRI dataset \cite{dataset1,dataset2} used in this study showing images containing three types of brain tumors (Meningioma, Glioma and Pituitary) in axial, coronal and sagittal views (see red arrows). This is a very challenging dataset where the tumor sizes vary across the population and tumor intensities and textures range from large bright areas with contrast surroundings to small dark areas with barely perceivable outlines.}
  \label{fig:dataset_sample}
\end{figure}
%\cite{dataset1,dataset2} This is a very challenging dataset where the tumor sizes vary across the population and tumor intensities and textures range from large bright areas with contrast surroundings to small dark areas with barely perceivable outlines.

There are several MRI datasets that are used for evaluation of deep learning algorithms to advance the classification performance. We focus on one specific and challenging 2D T1-weighted CE-MRI dataset \cite{dataset1}, compiled by Cheng et al. A random sample from this dataset is provided in Fig.\ref{fig:dataset_sample}. Many state-of-the-art classification algorithms (e.g. \cite{CNN_2022_2, CNN_2019, CNN_2022}) that produce good performance on this dataset are computationally expensive, especially at inference time. Addressing this limitation, our main contributions are: 
\begin{itemize}
  \item First, we develop a novel low-complexity Convolutional Neural Network (CNN) for classification of brain tumors and show that our proposed network outperforms all lightweight methods ($<$20ms in inference time) investigated on the Cheng et al. dataset.
 
\item Second, we design a new \textit{normalized} attention mechanism that can be added to any convolutional classification network to yield performance gains. 

  \item Third, we verify that the introduced $l_{2}$-normalization acts as a regularizer against overfitting during training and improves model convergence.
\end{itemize}

\section{Related Works and Motivation}
\label{sec:related_works}
In the realm of biomedical imaging, there have been many previous studies dedicated specifically to brain tumor classification. In particular, the MRI images from the Cheng et al. dataset \cite{dataset1} have been frequently used in recent studies.% as they include both segmentation masks and classification labels.% for the different types of tumors. 

Most recently, Aurna et al. \cite{CNN_2022_2} proposed a classification approach using a combination of feature extractors, optimizers, hyper-parameters, and PCA component reduction. Their method utilized a two-stage ensemble of ResNet-50 with a custom CNN and achieved 98.67\% classification accuracy.

Badža and Barjaktarovic \cite{CNN_2020} suggested classification via a custom CNN containing two convolutional blocks and an output block with two fully connected (FC) layers. They preprocessed the images with normalization, resizing, and augmentation, and reported an accuracy of 97.39\%.

Deepak and Ameer \cite{CNN_2019} preprocessed images by normalizing intensity using the min-max technique, then pretrained the deep network GoogLeNet with a custom modification. Their method of deep transfer learning reached an accuracy of 92.30\% using a standalone CNN.

Alanazi et al. \cite{CNN_2022} created an isolated, 22-layer CNN from scratch to differentiate binary classes, then applied a transfer-learning approach to differentiate tumor types. Their transfer-learned CNN model achieved 96.90\% accuracy.

Many other studies have explored the use of attention mechanisms as a method for improving the saliency of desired features in medical image processing. Woo et al.’s \cite{CBAM} Convolutional Block Attention Module (CBAM) proposed the use of both spatial and channel attention sub-modules in CNNs to  boost representation power, and outperformed the state-of-the-art Squeeze and Excitation \cite{SE} method. 

A vast majority of the existing research with attention mechanisms focuses on its use in image segmentation, however a few studies have recently explored the value of attention networks in classification. Schlemper et al. \cite{Schlemper} used additive attention with the feed-forward CNN model Sononet (AG-Sononet-16) and reported a maximum accuracy of 98\%. Xiao et al. \cite{TReC} presented a new approach specifically for classification of MRI small-scale samples. The model consisted of a transferred, pre-trained ResNet with custom FC layers and integrated CBAM into every residual block of ResNet. An average accuracy of 97.44\% was reported for multi-class classification with Transferred ResNet34-CBAM.

All models (except \cite{CNN_2020}) reviewed above have high computational burden in terms of the number of model parameters, training and inference time, and therefore are challenging to fit into the edge devices for medical applications. %This fact is the main motivation for our work. 
In our experiments we focus on obtaining performance enhancements compared to ``lightweight'' models, which we define as models that require $<20$ms for inference.

\section{Architecture}
\label{sec:architecture}

\begin{figure*}
  \centering
  \includegraphics[width=\linewidth]{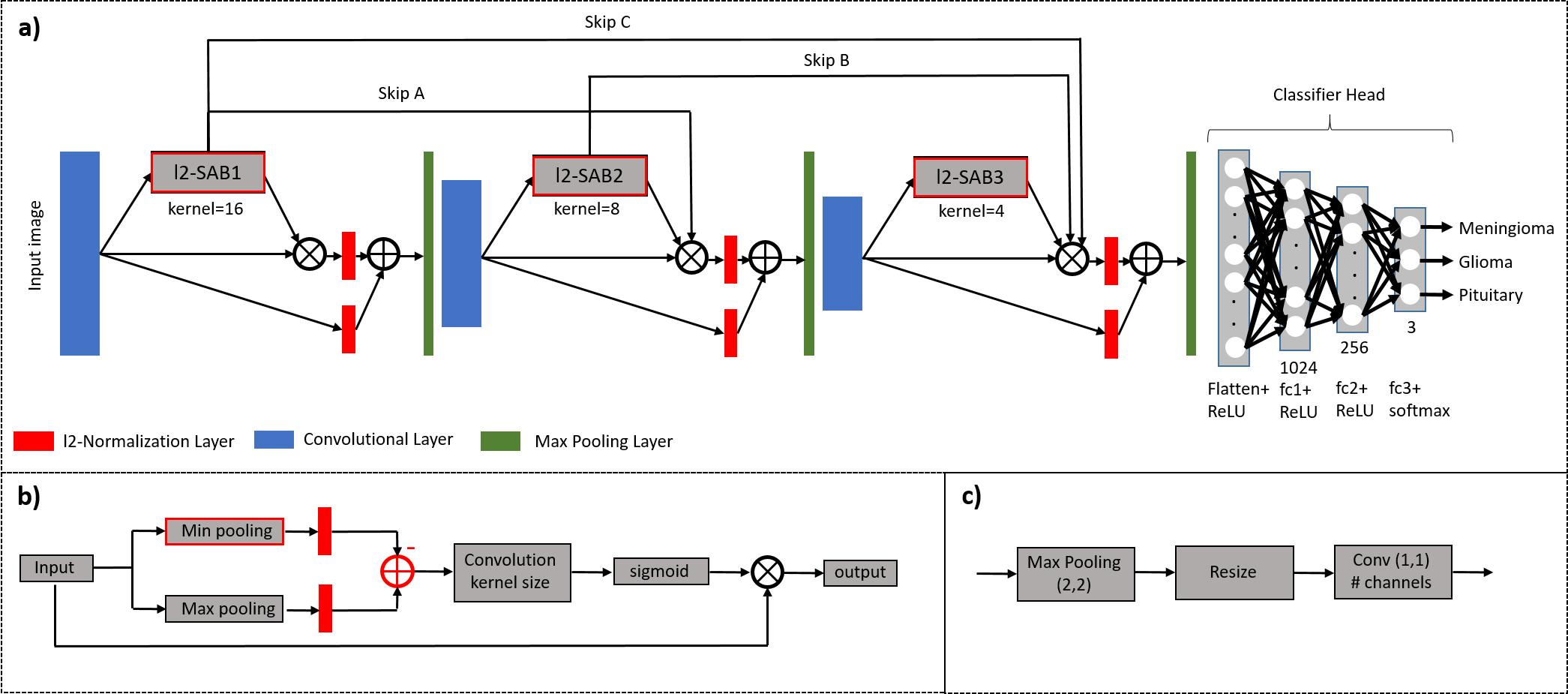}
  \caption{Overview of the proposed architecture. \textbf{a)} Proposed $l_{2}$-SA model with $l_{2}$-normalized spatial attention blocks ($l_{2}$-SAB). Shallow convolutional backbone (baseline) consists of only three Convolutional and three MaxPooling layers. The first MaxPooling layer is (4,4) and the remaining two are (2,2). The convolutional kernel sizes in the backbone decrease with the depth of the network (25-13-9) and the number of channels increases with the depth (64-128-256). \textbf{b)} Proposed $l_{2}$-normalized spatial attention block with our new additions in red. \textbf{c)} Inner organization of the multiplicative skip connections A, B and C.}
  \label{fig:architecture}
\end{figure*}

Our architecture consists of the convolutional backbone with only three Convolutional and three MaxPooling layers, and novel $l_{2}$-normalized spatial attention blocks ($l_{2}$-SAB) as shown in Fig.\ref{fig:architecture}. For the attention mechanism, we build upon the idea of CBAM first presented in \cite{CBAM}. The CBAM based attention mechanism amplifies the important features and suppresses the unimportant ones. In medical image analysis, CBAM-inspired attention has been e.g. used in \cite{CBAM_related_seg2022} and \cite{CBAM_related_seg2020} for the segmentation task and in \cite{CBAM_related_classify} for the classification of brain diseases. It is common to combine CBAM with the ResNet network \cite{CBAM_related_classify, CBAM}, but to our best knowledge, we are the first to integrate CBAM-like spatial attention into a shallow baseline model for fast classification of brain tumors in T1-weighted CE-MRI images. 

In this work, we propose a normalized spatial attention mechanism and call our new model $l_{2}$-SA. The introduced $l_{2}$-normalization layer normalizes a batch of inputs so that each input in the batch has a $l_{2}$ norm equal to 1. We empirically discovered that integrating the original CBAM framework into a fast baseline model does not yield any substantial performance gains on the classification task with the Cheng et al. dataset. We only  obtain a considerable accuracy gain and a model convergence after we introduce the $l_{2}$-normalization prior to the addition operation and include the modification of the spatial attention block (as illustrated in Fig.\ref{fig:architecture} a) and b)) . 

Our normalized spatial attention block differs from the spatial attention module of CBAM in two ways, as shown in Fig.\ref{fig:architecture} b) in red. First, we combine max and min feature maps instead of max and average features. Second, instead of concatenating the feature maps, we $l_{2}$-normalize and then subtract the min from the max features. The intuition behind this design is to increase the learning ability of the spatial attention block through a more discriminative feature combination \cite{MinMax_CNNs} by taking the difference between the informative max and min pooled features. We verify that this effective combination increases the representation power of the network. % and has only marginal computational overhead.

The following notations are adapted from \cite{CBAM}. Given a spatial attention map $\mathbf{M}_s \in {\rm I\!R}^{1 \times H \times W}$ and an intermediate feature map $\mathbf{F} \in {\rm I\!R}^{C \times H \times W}$, the output $\mathbf{F}'$ of the $l_{2}$-SAB can be written as (applied to each channel separately):
\begin{equation}
\mathbf{F}' = \mathbf{M}_s(\mathbf{F}) \otimes \mathbf{F}    
\end{equation}
where $\otimes$ denotes element-wise multiplication. Further, we use two pooling operations, $\mathbf{F}_{max}^s \in {\rm I\!R}^{1 \times H \times W}$ and $\mathbf{F}_{min}^s \in {\rm I\!R}^{1 \times H \times W}$. These operations produce two 2D feature maps that are further $l_{2}$-normalized, subtracted from each other and convolved by a convolutional layer with kernel size $K$:
\begin{equation}
\mathbf{M}_s(\mathbf{F}) = \sigma(f^{K \times K}(l_2(MaxPool(\mathbf{F}))-l_2(MinPool(\mathbf{F}))))    
\end{equation}
where $f^{K \times K}$ is a convolution operation with the filter size $K \times K$ and $\sigma$ denotes the sigmoid activation function.

\section{Methodology}
\label{sec:methodology}

\subsection{Preprocessing}
\label{subsec:preprocessing}

The images from the original Cheng et al. dataset were provided in different resolutions. In our data preparation step we save the grayscale images as pseudo-RGB with three stacked channels and resize to the input resolution of $256 \times 256 \times 3$. We do not apply any data augmentation. We do not perform any other preprocessing apart from rescaling by the value $255$.%Because the images are already in the range of $[0,1]$, we do not re-scale the intensity.  

\subsection{Dataset}
\label{subsec:dataset}

The 2D T1-weighted CE-MRI dataset\footnote{https://figshare.com/articles/dataset/brain\_tumor\_dataset/1512427} used in this work,  first mentioned in a 2015 study by Cheng et al, is publicly available \cite{dataset1}. The dataset was acquired from Nanfang Hospital, Guangzhou, China, and General Hospital, Tianjing Medical University, China, from 2005 to 2010. It is a  large and very challenging dataset as the same types of tumors have different appearances \cite{dataset2}. The total number of contrast enhanced images is 3064. The dataset has been collected from 233 patients and contains three categories of brain tumors: 708 Meningiomas, 1426 Gliomas and 930 Pituitary tumors in axial, coronal and sagittal views \cite{CNN_2022_2}. Therefore, the dataset is imbalanced as the prevalent category is Glioma. We randomly sample $2451$ images for training and validation and further reserve 2145 images for training and 306 images for validation. The test set contains 613 images. This represents a 70\% training, 10\% validation and 20\% test data split. %In order to mitigate the class imbalance, we compute the class weights that are proportional to the underrepresentation of classes.
%\begin{table}[h]
%\centering
%  \begin{tabular}{lllll}
%    \hline
%    Tumor Category & Train & Validation & Test & Weights \\
%    \hline
%    Meningioma & 461 & 102 & ...& 1.45 \\
%    Glioma & 928 & 207  & ...   & 0.72 \\
%    Petituary & 611 & 142 & ... & 1.09 \\
%    \hline
%    Total & 2000 & 451 & 613&\\
%    \hline
%  \end{tabular}
%  \caption{Statistics of the Cheng et al. dataset showing the number of images after the random split of the original dataset. It is evident that it is an imbalanced dataset with the dominant category of Glioma.}
%  \label{tab:dataset_stats}
%\end{table}

\subsection{Implementation details}
\label{subsec:implementation}

The classification pipeline was implemented in Python 3.9.12, Tensorflow 2.9.1 and Keras 2.9.0. All experiments were performed on a desktop computer with the Ubuntu operating system 18.04.3 LTS with the Intel(R) Core(TM) i9-9900K CPU and a total of 62GB RAM. 

We train the proposed $l_{2}$-SA model for 50 epochs using Adam optimizer with  learning rate of 0.01, epsilon=0.1 and the batch size of 64. We use sparse categorical crossentropy as the loss function.% because the data are in folders. 

\begin{table*}[t!]
\centering
  \begin{tabular}{lllllll}
    \hline
     \textbf{Classification Model}  & \textbf{Year} & \textbf{Result from:} & \textbf{Accuracy}  & \textbf{Evaluation} & \textbf{\# parameters} & \textbf{Inf.}\\
     \hline
    BoW+classifier \cite{dataset1} &2015 & Original paper & 91.28\% & n/a & n/a &  n/a \\
    %\hline
    CNN \cite{CNN_2019} Transfer Learning & 2019 &  Original paper & 93.0\%   & five-fold & -  & - \\
    %\hline
    CNN \cite{CNN_2020} & 2020 &  Reproduced & 95.27\%  & 70\% - 10\% - 20\% & 1,282,955 & 10ms  \\
    %\hline
    CNN \cite{CNN_2022} Transfer Learning & 2022 & Original paper & 95.75\%  & 80\% - 20\% &  - & -  \\
    %\hline
    CNN Ensemble \cite{CNN_2022_2} & 2022 & Original paper & 98.67\%  & 70\% - 30\% & -  & -  \\
    %\hline
    VGG16* & &  Reproduced & 94.29\% & 70\% - 10\% - 20\% & 25,972,491 & 18ms \\
    %\hline
    baseline  & & Reproduced & 94.78\% &  70\% - 10\% - 20\% &4,003,011 & 16ms\\
    %\hline
    baseline + CBAM \cite{CBAM}  && Reproduced & 95.60\% &  70\% - 10\% - 20\% & 5,474,547  & 20ms\\
    %\hline
    Fine-tuned VGG16  && Reproduced & 93.31\% &  70\% - 10\% - 20\% &  114,239,171 & 109ms\\
    %(pre-trained on ImageNet) & &&&&&\\
    %\hline
    \textbf{Proposed $l_{2}$-SA}  && Reproduced & \textbf{96.57\%} &  70\% - 10\% - 20\% & 7,293,523 & \textbf{19ms}\\
    Proposed $l_{2}$-SA + VGG16 ensemble  && Reproduced & 96.74\% &  70\% - 10\% - 20\% & 15,516,179 & 115ms\\
    Proposed $l_{2}$-SAB + CNN \cite{CNN_2020}  && Reproduced & 96.08\% &  70\% - 10\% - 20\% & 2,199,811  & 13ms\\
    \hline
  \end{tabular}
  \caption{Quantitative evaluation on the Cheng et al. dataset and comparison with the state-of-the-art. Best results are shown. The best accuracy gain of 1.79 percentage points is obtained when comparing the baseline model (shallow convolutional backbone + classifier head) with the proposed $l_{2}$-SA model. An even better result is achieved through the VGG16 ensemble at the expense of increased inference time. Both VGG16 models (fine-tuned and ensemble) are pre-trained on the ImageNet dataset. Adding the proposed $l_{2}$-SAB on top of the fast CNN \cite{CNN_2020} results in the improved 96.08\%. This shows that our $l_{2}$-normalized spatial attention mechanism can benefit any convolutional model and has only 3ms computational overhead.}
  \label{tab:class_performance}
\end{table*}
%A fast CNN model from \cite{CNN_2020} is our implementation, and we can not reproduce the 97.39\% accuracy reported by the authors. Finally, we outperform all lightweight models reviewed in the category of fast inference time $< 20$ms.
\subsection{Experiments}
\label{subsec:experiments}
Training for each model is repeated 20 times and the best result is reported. Comparison with the state-of-the-art is provided in Table~\ref{tab:class_performance}. Our proposed $l_{2}$-SA model reaches  $96.57$\% accuracy and outperforms all lighweight models investigated in the category of fast inference time.%, i.e. $< 20$ms.
%\begin{figure}[h]
%  \centering
%  \vspace{0pt}
%  \includegraphics[width=7cm]{conf_matrix_9657acc.png} 
%  \caption{Confusion matrix for the best result with the classification %accuracy = 96.57\% of the proposed model ANSA.}
%  \label{fig:conf_matrix}
%\end{figure}

For benchmarking, we build several networks from scratch including VGG16*. This model consists of three convolutional blocks (each with two Convolutional layers) and three (2,2) MaxPooling layers. Convolutional layers have increasing number of channels (8-16-32), and the constant kernel sizes of (3-3-3). The classifier head is the same as with the baseline and the $l_{2}$-SA models, and is detailed in Fig.~\ref{fig:architecture}. %Experiments show that the classification accuracy of the somewhat slower VGG16* lags behind that of the AlexNet*.

We also perform an experiment where we integrate the spatial attention module from the original CBAM paper \cite{CBAM} into the baseline model. During training, repeated 20 times, there are cases where this composite model fails to converge or overfits. The best result of this experiment shows an improvement of 0.82 percentage points versus the baseline.% model.

In order to reproduce the results from \cite{CNN_2020}, we build this fast CNN according to the details provided in the original paper. However, we can not reproduce the best result reported in \cite{CNN_2020}. Further, we remove the Dropout layers and add the proposed $l_{2}$-SAB blocks with the kernels (16-8-4-4) to this model. We observe that the addition of $l_{2}$-SAB blocks results in improved performance of 96.08\%. We think that even better performance can be reached by tuning the kernels of the $l_{2}$-SAB blocks. %This could e.g. be accomplished with the Keras Tuner if the according computational resources are available.

The performances of the fine-tuned VGG16 model and that of the combination of $l_{2}$-SA and the VGG16 ensemble demonstrate that the inference time of comparable transfer learning based models lies in the region of $> 100$ms. 

Generally, reviewing the state-of-the-art and drawing conclusions over the leadership on this benchmark proved to be non-trivial. There are a number of reasons for this. First, none of the published papers provided either the code or the random data split indices to reproduce the results. Second, different papers use different evaluation protocols such as five-fold, ten-fold, subject-based and random data splits. Third, many papers do not document the inference time.% and hardware specifications.% into training, validation and test data. %Many papers reviewed do not provide details about the inference time, leaving the reader to estimate the time based on the architectural concepts provided. 

\subsection{Ablation study}
\label{subsec:ablation}
In order to demonstrate the contribution of  the skip connections in the $l_{2}$-SA model, we conduct an ablation study. Without the skips the best accuracy is 96.41\%. After the introduction of the A, B and C skip connections the best accuracy increases to 96.57\%. Also, we confirm that the ablation of the $l_{2}$-normalization layers results in the model overfitting.

%\section{Copyright forms}
%\label{sec:copyright}

%You must include your fully completed, signed IEEE copyright release form when you submit your paper. We {\bf must} have this form before your paper can be published in the proceedings.  The copyright form is available as a Word file, a PDF file, and an HTML file. You can also use the form sent with your author kit.

\section{Conclusion}
\label{sec:conclusion}

In this work, we propose a normalized spatial attention mechanism and show that when combined with a baseline classification network this results in a significant accuracy gain of 1.79 percentage points in the targeted classification task. Furthermore, we outperform all lightweight models investigated and are competitive with the computationally expensive methods involving transfer learning and ensembles of pre-trained CNNs. We envisage that given its speed and accuracy our approach will be of particular interest to edge computing for medical applications. Future work will focus on conducting experiments on different MRI datasets to test the generality of the proposed approach to image classification.

\section{Acknowledgments}
\label{sec:acknowledgments}
This research is partially supported by the NSF OISE program
award 2107439.

%\section{Compliance with ethical standards}
%This research study was conducted retrospectively using human subject data made available in open access. Ethical approval was *not* required as confirmed by the license attached with the open access data.
\vfill\pagebreak

%\section{REFERENCES}
%\label{sec:refs}

\bibliographystyle{IEEEbib}
\bibliography{strings,refs}

\end{document}